\documentstyle[preprint,aps,prb,tighten]{revtex}
\begin{document}
\draft
\title{Semiclassical theory of shot-noise suppression}

\author{M. J. M. de Jong$^{a,b}$ and C. W. J. Beenakker$^{b}$}

\address{
(a) Philips Research Laboratories,
5656 AA  Eindhoven,
The Netherlands\\
(b) Instituut-Lorentz,
University of Leiden,
2300 RA  Leiden,
The Netherlands
}

\date{submitted December 13, 1994 --- {\tt cond-mat/9501004}}

\maketitle

\begin{abstract}
The Boltzmann-Langevin equation is used to relate the shot-noise
power of a mesoscopic conductor to classical transmission
probabilities at the Fermi level.
This semiclassical theory is applied to tunneling through $n$
barriers in series.
For $n \rightarrow \infty$
the shot noise approaches one third
of the Poisson noise, independent of the transparency of the barriers.
This confirms that the one-third suppression known
to occur in diffusive conductors does not require phase coherence.
\end{abstract}

\pacs{PACS numbers: 72.70.+m, 73.50.Td, 72.10.Bg}

\narrowtext

The discreteness of the electron charge causes time-dependent
fluctuations in the electrical current,
known as shot noise. These fluctuations are characterized by a white
noise spectrum and persist down to zero temperature.
The shot-noise power $P$ contains information on the conduction process
which is not given by the resistance.
A well-known example is a vacuum diode, where
$P=2 e |I|\equiv P_{\text{Poisson}}$, with $I$ the average current.
This tells us that the electrons traverse the conductor in completely
uncorrelated fashion, as in a Poisson process.
In macroscopic samples the shot noise is averaged out to zero by
inelastic scattering.

In the last few years, the shot noise has been investigated
in mesoscopic conductors, smaller than the
inelastic scattering length. Theoretical analysis
shows that the shot noise can be
suppressed below $P_{\text{Poisson}}$, due to correlations in the electron
transmission imposed by the Pauli
principle.\cite{khl87,les89,yur90,but90,mar92}
Most intriguingly, it has been found that
$P=\case{1}{3} P_{\text{Poisson}}$
in a metallic, diffusive
conductor.\cite{b&b92,nag92,jon92,naz94,alt94}
The factor one third is universal in the sense that it is independent
of the material, sample size, or degree of disorder,
as long as the length $L$ of the conductor is greater than
the mean free path $\ell$ and shorter than the localization length.
An observation of
suppressed shot noise in a diffusive conductor has been
reported.\cite{lie94}
In a quantum-mechanical description\cite{b&b92}
the suppression follows from the
bimodal distribution of transmission eigenvalues.\cite{dor84}
Surprisingly, Nagaev\cite{nag92} finds the same one-third suppression
from a semiclassical approach, in which the Pauli principle is accounted
for, but
the motion of electrons
is treated classically.
This implies that
phase coherence is not essential for the suppression, but the relationship
between the quantum-mechanical and
semiclassical theories remains unclear.

In this paper we reinvestigate the semiclassical
approach and present a detailed comparison with quantum-mechanical
calculations in the literature.
We use the
Boltzmann-Langevin equation,\cite{kad57,kog69}
which is a semiclassical kinetic equation
for non-equilibrium fluctuations.
This equation has previously been applied to shot noise
by Kulik and Omel'yanchuk\cite{kul84}
for a ballistic point contact,
and by Nagaev\cite{nag92} for a diffusive conductor.
Below we will demonstrate how the
Boltzmann-Langevin equation can be applied to
an arbitrary mesoscopic conductor.
Our analysis corrects a previous paper.\cite{bee91}
To be specific,
we consider tunneling through $n$ planar barriers in series
(tunnel probability $\Gamma$).
This model is sufficiently simple that it allows us to obtain a closed
expression for $P$ and sufficiently general that we can compare with all
results in the literature from the quantum-mechanical approach.
For $n=2$ and $\Gamma \ll 1$ we recover the results for a
double-barrier junction of Refs.\ \onlinecite{che91}
and \onlinecite{dav92}.
In the limit $n \rightarrow \infty$ the shot-noise power approaches
$\case{1}{3} P_{\text{Poisson}}$ independent of $\Gamma$.
By taking the continuum limit, $n \rightarrow \infty$,
$\Gamma \rightarrow 1$, at fixed $n (1-\Gamma)$,
we can study the shot noise in a diffusive conductor
and the crossover to ballistic transport. We find
exact agreement with a previous
quantum-mechanical evaluation,\cite{jon92}
in the limit of a conductance $\gg e^2/h$.
It has been emphasized by Landauer,\cite{lan94}
that Coulomb interactions may induce a further reduction
of $P$.
Here we follow the quantum-mechanical
treatments in assuming non-interacting electrons,
leaving interaction effects for future work.

Let us begin by briefly reviewing the quantum-mechanical approach.
The zero-temperature, zero-frequency shot-noise power $P$
of a phase-coherent conductor is related to the
transmission matrix $\underline{t}$ by the formula\cite{les89,but90}
\begin{equation}
P = P_0 \mbox{Tr} \,
 \underline{t} \, \underline{t}^\dagger
( \underline{1} - \underline{t} \,
\underline{t}^\dagger )
= P_0 {\textstyle \sum_{n=1}^{N}}
T_n ( 1 - T_n ) \: ,
\label{e1}
\end{equation}
where $P_0\equiv 2 e |V| G_0$,
with $V$ the applied voltage and $G_0 \equiv e^2/h$ the conductance
quantum (we assume spinless electrons for simplicity of notation),
$T_n \in [0,1]$ an eigenvalue of
$\underline{t} \, \underline{t}^\dagger$, and $N$ the number of transverse
modes at the Fermi energy $E_F$.
The conductance is given by the Landauer formula
\begin{equation}
G = G_0 \mbox{Tr} \, \underline{t} \, \underline{t}^\dagger
 = G_0 {\textstyle \sum_{n=1}^{N}} T_n
\: .
\label{e2}
\end{equation}
One finds $P=2 e |V| G= P_{\text{Poisson}}$
for a conductor where all
$T_n \ll 1$ (such as a high tunnel barrier).
However, if some $T_n$ are near 1 (open channels), then the shot noise is
reduced below $P_{\text{Poisson}}$.
In a phase-coherent metallic, diffusive conductor the $T_n$ are either
exponentially small or of order unity.\cite{dor84}
This bimodal distribution gives rise to the one-third suppression of the
shot noise.\cite{b&b92}

We now formulate the semiclassical kinetic theory.\cite{kad57,kog69}
We consider a $d$-dimensional conductor
connected by ideal leads to two electron reservoirs.
The {\em fluctuating} distribution function
$f({\bf r}, {\bf k}, t)$ equals $(2 \pi)^d$ times the density
of electrons with position ${\bf r}$,
and wave vector ${\bf k}$, at time $t$.
(The factor $(2 \pi)^d$ is introduced so that $f$ is the occupation number
of a unit cell in phase space.)
The average over time-dependent fluctuations
$\langle f \rangle \equiv \bar{f}$
obeys the Boltzmann equation
\begin{mathletters}
\label{e3}
\begin{eqnarray}
&&\left( d/dt + {\cal S} \right)
\bar{f}({\bf r}, {\bf k}, t) = 0 \: ,
\label{e3a}
\\
&&\frac{d}{dt} \equiv
\frac{\partial}{\partial t} +
{\bf v} \cdot \frac{\partial}{\partial {\bf r}} +
{\cal F} \cdot \frac{\partial}{\hbar \partial {\bf k}} \: .
\label{e3b}
\end{eqnarray}
\end{mathletters}%
The derivative (\ref{e3b})
(with ${\bf v}=\hbar {\bf k}/m$)
describes the classical motion in the force field ${\cal F}({\bf r})$.
The term ${\cal S} \bar{f}$ accounts for the
stochastic effects of scattering.
Only elastic scattering is taken into account
and electron-electron scattering is disregarded.
We consider the stationary situation where
$\bar{f}$ is independent of $t$.
The time-dependent fluctuations
$\delta f \equiv f - \bar{f}$ satisfy
the Boltzmann-Langevin equation\cite{kad57,kog69}
\begin{equation}
\left( d/dt + {\cal S} \right)
\delta f ({\bf r}, {\bf k}, t) =
j ({\bf r}, {\bf k}, t) \: ,
\label{e5}
\end{equation}
where $j$ is a fluctuating source term.
In the Boltzmann equation (\ref{e3}) scattering occurs into all wave
vectors $\bf k$ with some probability distribution.
Eq.\ (\ref{e5}) takes into account that
each electron is scattered into only one particular $\bf k$.
This implies that the flux $j$
is positive for that $\bf k$ and negative for the others.
The flux $j$ has zero average,
$\langle j \rangle = 0$, and covariance
\begin{equation}
\langle j ({\bf r}, {\bf k}, t) j ({\bf r}', {\bf k}', t') \rangle
=
(2 \pi)^d \delta({\bf r} -{\bf r'}) \delta(t-t')
J({\bf r}, {\bf k}, {\bf k}') \: .
\label{e6}
\end{equation}
The delta functions ensure that
fluxes are only correlated
if they are induced by the same scattering process.
The correlator $J$ depends on the type of scattering and on $\bar{f}$,
but not on $\delta f$.
The $J$ for impurity scattering is given in Ref.\ \onlinecite{kog69},
and for barrier scattering it is given below.

We evaluate the fluctuating current $\delta I(t) \equiv I(t) - \bar{I}$
through cross-section $S_r$ in the right lead,
\begin{equation}
\delta I (t) = \frac{e}{(2 \pi)^d}
\int_{S_r} \! \! d{\bf y} \int \! d {\bf k} \:
v_x \, \delta f ({\bf r}, {\bf k}, t) \: .
\label{e7}
\end{equation}
Here ${\bf r}=(x,{\bf y})$,
with the $x$-coordinate along and ${\bf y}$
perpendicular to the lead.
The zero-frequency noise power
is given by
\begin{equation}
P \equiv 2 \int_{-\infty}^{\infty} \! \!
dt \, \langle \delta I (t) \delta I (0) \rangle \: .
\label{e8}
\end{equation}
The formal solution of Eq.\
(\ref{e5}) is
\begin{eqnarray}
\delta f ({\bf r}, {\bf k}, t)&=&
\int_{-\infty}^{t} \! \! dt' \int \! d{\bf r}' \int \! d{\bf k}'
\nonumber \\ &&
{\cal G}( {\bf r}, {\bf k}; {\bf r}', {\bf k}'; t-t') \,
j({\bf r}', {\bf k}', t') \: ,
\label{e9}
\end{eqnarray}
where the Green's function $\cal G$ is a solution of
\begin{equation}
\left( d/dt + {\cal S} \right)
{\cal G}( {\bf r}, {\bf k}; {\bf r}', {\bf k}';t)
= \delta({\bf r}-{\bf r}')
\delta({\bf k}-{\bf k}') \delta(t) \: ,
\label{e10}
\end{equation}
such that ${\cal G}=0$ if $t < 0$.
The transmission probability $T({\bf r}, {\bf k})$
is the probability that an electron at $({\bf r}, {\bf k})$
leaves the wire through the right lead.
It is related to ${\cal G}$ by
\begin{equation}
T({\bf r}, {\bf k}) =
\int_0^{\infty} \!\! dt
\int_{S_r} \!\! d{\bf y}'
\int \! d{\bf k}' \, v_x' \,
{\cal G}( {\bf r}', {\bf k}'; {\bf r}, {\bf k};t) \: .
\label{e11}
\end{equation}
Eqs.\ (\ref{e6})--(\ref{e11}) yield for the noise power the expression
\begin{equation}
P=\frac{2 e^2}{ (2 \pi)^d} \int \! \! d{\bf r} \! \!
\int \! \! d {\bf k} \! \! \int \! \! d {\bf k}' \,
T({\bf r},{\bf k}) T({\bf r},{\bf k}')
J({\bf r},{\bf k},{\bf k}') \: .
\label{e12}
\end{equation}

Eq.\ (\ref{e12}) applies generally to any conductor in which only elastic
scattering occurs. We now specialize to the case
that the scattering is due to
$n$ planar tunnel
barriers, perpendicular to the $x$-direction
(see inset of Fig.\ \ref{f1}).
Barrier $i$ has tunnel probability $\Gamma_i \in [0,1]$,
which for simplicity
is assumed to be $\bf k$ and ${\bf y}$-independent.
Upon transmission $\bf k$ is conserved, whereas upon reflection
${\bf k} \rightarrow \widetilde{\bf k}\equiv(-k_x,{\bf k}_y)$.
In what follows we drop the (irrelevant) transverse coordinate $\bf y$.
At barrier $i$ (at $x=x_i$)
the average densities $\bar{f}$
on the left side ($x_{i-}$) and on the
right side ($x_{i+}$) are related by
\begin{equation}
\bar{f}(x_{i+},{\bf k}) = \Gamma_i \bar{f}(x_{i-},{\bf k})
+ (1 - \Gamma_i)  \bar{f}(x_{i+},\widetilde{\bf k}) \: ,
\label{e13}
\end{equation}
for $k_x > 0$.
The relation for $k_x<0$ is Eq.\ (\ref{e13}) with $x_{i-}$ and
$x_{i+}$ interchanged.
To determine the correlator $J$ in Eq.\ (\ref{e6}) we argue in a similar
way as in Ref.\ \onlinecite{mar92}.
Consider an incoming state from the left $(x_{i-},{\bf k})$
and from the right $(x_{i+},\widetilde{\bf k})$
(we assume $k_x > 0$). If both incoming states
are either filled or empty, there will be no fluctuations
in the outgoing states, hence $j=0$.
Let us therefore consider the case that the incoming
state from the left is filled and that from the right is empty, which
occurs with probability
$\bar{f}(x_{i-},{\bf k}) [ 1 - \bar{f}(x_{i+},\widetilde{\bf k}) ]$.
On the average, the outgoing states at the left and right
have occupation $1-\Gamma_i$ and $\Gamma_i$, respectively.
However, since the incoming electron is either
transmitted or reflected, the instantaneous occupation of the
outgoing states differs from the average occupation. Upon transmission,
the state at the right (left) has an excess (deficit) occupation of
$1-\Gamma_i$.
Transmission occurs with probability $\Gamma_i$, so the
contribution to $J(x,{\bf k},{\bf k}')$ from a transmitted electron is
$\Gamma_i(1-\Gamma_i)^2
[\delta({\bf k}-{\bf k}')-\delta({\bf{k}}-\widetilde{\bf k}')]
\delta(x-x_i)|v_x|$.
Similarly, a reflected electron contributes
$(1-\Gamma_i) \Gamma_i^2
[\delta({\bf k}-{\bf k}')- \delta({\bf k}-\widetilde{\bf k}')]
\delta(x-x_i)|v_x|$.
Collecting results, we find for $k_x>0$
\widetext
\begin{eqnarray}
J(x,{\bf k},{\bf k}')&=&{\textstyle \sum_{i=1}^n}
\delta(x-x_i) \, \Gamma_i (1-\Gamma_i) \, |v_x|
[\, \delta({\bf k}-{\bf k}') - \delta({\bf k}-\widetilde{\bf k}') \, ]
\times
\nonumber \\
&&
\Bigl\{ \bar{f}(x_{i-},{\bf k}) [\, 1 - \bar{f}(x_{i+},\widetilde{\bf k}) \, ]
+
\bar{f}(x_{i+},\widetilde{\bf k}) [\, 1 - \bar{f}(x_{i-},{\bf k}) \, ] \Bigr\}
\: .
\label{e14}
\end{eqnarray}
\narrowtext%
For $k_x < 0$, Eq.\ (\ref{e14}) holds upon interchanging
$x_{i-}$ and
$x_{i+}$.

The average distribution function $\bar{f}$ inside the conductor
depends on the equilibrium distributions $f_l$ and $f_r$
in the left and right reservoirs, according to
\begin{equation}
\bar{f}(x, {\bf k}) =
T(x, -{\bf k}) f_r(\varepsilon)
+ [1- T(x, -{\bf k})] f_l(\varepsilon) \: ,
\label{e15}
\end{equation}
where $\varepsilon$ is the electron energy and
$T(x, -{\bf k})$ equals the probability that an electron
at $(x, {\bf k})$ has arrived there from the right reservoir.
At zero temperature one has
$f_l(\varepsilon )=\Theta(E_F + e V - \varepsilon)$,
$f_r(\varepsilon )=\Theta(E_F - \varepsilon)$,
with $\Theta$ the unit step-function.
Substitution of Eqs.\ (\ref{e14}) and (\ref{e15})
into Eq.\ (\ref{e12}) and linearization in $V$ yields
\begin{eqnarray}
P &=& P_0 N \,
{\textstyle \sum_{i=1}^n} \Gamma_i (1-\Gamma_i) \times
\nonumber \\ &&
(T_i^\rightarrow-T_i^\leftarrow)^2 (T_i^\rightarrow +
T_i^\leftarrow - 2 T_i^\rightarrow T_i^\leftarrow )
\: ,
\label{e16}
\end{eqnarray}
where $T_i^\rightarrow\equiv T(x_{i+},k_x>0)$
[$T_i^\leftarrow\equiv T(x_{i-},k_x<0)$] is the transmission probability
into the right reservoir
of an electron moving away from the right [left] side of barrier $i$.
The conductance is given simply by
\begin{equation}
G=G_0 N \, T_0 \: ,
\label{e17}
\end{equation}
where $T_0\equiv T(x_{1-},k_x>0)$ is the
transmission probability through the whole conductor.
Comparing Eqs.\ (\ref{e2}) and (\ref{e17}) we note
that $\sum T_n$ corresponds semiclassically
to $N T_0$.
Comparison of Eqs.\ (\ref{e1}) and (\ref{e16}) shows that the semiclassical
correspondence to $\sum_n T_n(1-T_n)$ is much more complicated, as it
involves the transmission probabilities $T_i^\rightarrow,T_i^\leftarrow$
at all scatterers inside
the conductor (and not just the transmission probability $T_0$ through
the whole conductor).

As a first application of Eq.\ (\ref{e16}) we calculate the shot noise for
a single tunnel barrier. Using $T_0=\Gamma$,
$T_1^\leftarrow=0$, $T_1^\rightarrow=1$, we find
the expected result\cite{khl87,les89,yur90,but90,mar92}
$P=P_0 N \Gamma(1-\Gamma) = (1-\Gamma)
P_{\text{Poisson}}$.
The double-barrier case ($n=2$) is less trivial.
Experiments by Li {\em et al.},\cite{li90} showed full Poisson noise
for asymmetric structures ($\Gamma_1 \ll \Gamma_2$)
and a suppression by
one half for the symmetric case ($\Gamma_1 \simeq \Gamma_2$).
This effect has been explained by
Chen and Ting,\cite{che91} by Hershfield {\em et al.},\cite{dav92}
and by others.\cite{her92}
These theories assume resonant tunneling in the regime that the applied
voltage $V$ is much greater than the width of the resonance.
This requires
$\Gamma_1, \Gamma_2 \ll 1$. The present semiclassical
approach makes no reference to transmission resonances and
is valid for all
$\Gamma_1, \Gamma_2$.
For the double-barrier system one has
$T_0=\Gamma_1 \Gamma_2/\Delta$,
$T_1^\leftarrow=0$,
$T_1^\rightarrow=\Gamma_2/\Delta$,
$T_2^\leftarrow=(1-\Gamma_1)\Gamma_2/\Delta$, and
$T_2^\rightarrow=1$, with $\Delta=\Gamma_1+\Gamma_2 - \Gamma_1 \Gamma_2$.
 From Eqs.\ (\ref{e16}) and (\ref{e17}) it follows that
\begin{equation}
P =
\frac{\Gamma_1^2 (1-\Gamma_2) + \Gamma_2^2 (1-\Gamma_1)}
{(\Gamma_1+\Gamma_2 - \Gamma_1 \Gamma_2)^2} \, P_{\text{Poisson}}
\: .
\label{e19}
\end{equation}
In the limit $\Gamma_1, \Gamma_2 \ll 1$
Eq.\ (\ref{e19}) coincides precisely with the results of Refs.\
\onlinecite{che91} and \onlinecite{dav92}.

The shot-noise suppression of one half for a symmetric
double-barrier junction has the same origin
as the one-third suppression for a diffusive conductor.
In our semiclassical model, this is evident from the
fact that a diffusive conductor is the continuum limit
of a series of tunnel barriers. We demonstrate this below.
Quantum-mechanically, the common origin is the bimodal distribution
$\rho(T)\equiv \langle \sum_{n} \delta(T-T_n) \rangle$
of transmission eigenvalues, which for a double-barrier junction is
given by\cite{mel94}
\begin{equation}
\rho(T)=
\frac{N \Gamma_1 \Gamma_2}{\pi T
\sqrt{4 \Gamma_1 \Gamma_2 \, T - (\Delta \, T + \Gamma_1 \Gamma_2)^2}}
\: ,
\label{e20}
\end{equation}
for $T \in [T_-, T_+]$, with
$T_\pm=\Gamma_1 \Gamma_2 / (1 \mp \sqrt{1-\Delta} )^2$.
For a symmetric junction ($\Gamma_1=\Gamma_2\ll 1$),
the density (\ref{e20}) is
strongly peaked near $T=0$ and $T=1$, leading to a suppression of shot noise,
just as in the case of a diffusive conductor. In fact, one can verify that
the average of Eqs.\ (\ref{e1}) and (\ref{e2})
with the bimodal distribution (\ref{e20})
gives precisely the result (\ref{e19}) from the Boltzmann-Langevin equation.

We now consider $n$ barriers
with equal $\Gamma$.
We find
$T_0=\Gamma/\Delta$,
$T_i^\rightarrow=[\Gamma+ i(1-\Gamma)]/\Delta$,
and
$T_i^\leftarrow=(i-1)(1-\Gamma)/\Delta$, with
$\Delta=\Gamma+ n(1-\Gamma)$.
Substitution  into Eqs.\ (\ref{e16}) and (\ref{e17}) yields
\begin{equation}
P = \frac{1}{3}
\left( 1 +
\frac{n (1-\Gamma)^2 (2+\Gamma) - \Gamma^3}
{[\Gamma + n (1-\Gamma)]^3} \right) P_{\text{Poisson}}
\: .
\label{e22}
\end{equation}
The shot-noise suppression for a low barrier ($\Gamma=0.9$)
and for a high barrier ($\Gamma=0.1$) is plotted
against $n$ in Fig.\ \ref{f1}a.
For $\Gamma=0.1$ we observe almost full shot noise if $n=1$,
one-half suppression if $n=2$, and on increasing $n$
the suppression rapidly reaches one third.
For $\Gamma=0.9$ we observe that $P/P_{\text{Poisson}}$ increases
from almost zero to one third.
It is clear from Eq.\ (\ref{e22}) that
$P \rightarrow \case{1}{3} P_{\text{Poisson}}$
for $n \rightarrow \infty$
{\em independent} of $\Gamma$.

Finally, we make the connection with elastic impurity scattering
in a disordered wire.
Here, the scattering occurs throughout the whole wire instead of
at a discrete number of barriers.
We have previously carried out a quantum-mechanical study of
the shot noise in such a wire,\cite{jon92} on the basis
of the Dorokhov-Mello-Pereyra-Kumar equation.\cite{dor82}
For the semiclassical evaluation
we take the limit $n \rightarrow \infty$ and
$\Gamma \rightarrow 1$, such that
$n (1-\Gamma) = L/\ell$.
 From Eq.\ (\ref{e17}) one then obtains the
conductance $G=G_0 N (1+L/\ell)^{-1}$.
For the shot-noise power we find from Eq.\ (\ref{e22}),
\begin{equation}
P=\case{1}{3}
\left[ 1 -
(1 + L/\ell)^{-3} \right]  P_{\text{Poisson}}
\: .
\label{e23}
\end{equation}
This is precisely the result of Ref.\ \onlinecite{jon92}
for a metallic wire ($N \ell/L \gg 1)$.
Eq.\ (\ref{e23}) is plotted in Fig.\ \ref{f1}b and
describes how the shot noise
crosses over from complete suppression in the ballistic regime to
one third of the Poisson noise in the diffusive regime.
The diffusive limit confirms Nagaev's calculation.\cite{nag92}
Quantum corrections (of order $P_0$) to the shot-noise power
due to weak localization\cite{jon92} cannot be obtained within
our semiclassical approach.

In summary, we have presented a general framework
to derive the shot noise from the semiclassical
Boltzmann-Langevin equation, and applied this to the case of conduction
through a sequence of tunnel barriers.
We obtain a sub-Poissonian shot-noise power, in complete agreement with
quantum-mechanical calculations in the literature.
This establishes that phase coherence is not required for the
occurrence of suppressed shot noise in mesoscopic conductors.

This research was supported
by the Dutch Science Foundation NWO/FOM.

\begin{figure}
\vspace{0.5cm}
\caption{
(a) The shot-noise power $P$ for
$n$ tunnel barriers in series with transmission probability
$\Gamma=0.1$ (dots) and $\Gamma=0.9$ (circles), computed from Eq.\
(\protect\ref{e22}).
The dashed line is the large-$n$ limit $P=\case{1}{3} P_{\text{Poisson}}$.
The inset shows schematically the geometry considered.
(b) The shot-noise power $P$ of a disordered wire
as a function of the ratio of length $L$
to mean free path $\ell$, according to
Eq.\ (\protect\ref{e23}).}
\label{f1}
\end{figure}

\end{document}